\documentclass[amsmath,amssymb,aps,prl,twocolumn,longbibliography]{revtex4-1}

\usepackage{amssymb}
\usepackage{graphicx}
\usepackage{dcolumn}
\usepackage{bm}
\usepackage{amsmath}
\usepackage{mathrsfs}

\usepackage{textcomp}

\usepackage{wordlike}

\usepackage[unicode=true,bookmarks=true,bookmarksnumbered=false,bookmarksopen=false,breaklinks=false,pdfborder={0 0 1},backref=false,colorlinks=true]{hyperref}

\hypersetup{linkcolor=magenta,urlcolor=blue,citecolor=blue,pdfstartview={FitH},hyperfootnotes=false,unicode=true}

\def\be{\begin{equation}}
\def\ee{\end{equation}}
\def\bea{\begin{eqnarray}}
\def\eea{\end{eqnarray}}

\begin{document}
\title{Dynamic formation of quasicondensate and spontaneous vortices in a strongly interacting Fermi gas}
\author{Xiang-Pei Liu$^{1,2,3}$}
\thanks {These authors contributed equally to this work.}
\author{Xing-Can Yao$^{1,2,3}$}
\thanks {These authors contributed equally to this work.}
\author{Youjin Deng$^{1,2,3,6}$}
\thanks {These authors contributed equally to this work.}
\author{Yu-Xuan Wang$^{1,2,3}$}
\author{Xiao-Qiong Wang$^{1,2,3,}$}
\author{Xiaopeng Li$^{4,5}$}
\author{Qijin Chen$^{1,2,3}$}
\author{Yu-Ao Chen$^{1,2,3}$}
\author{Jian-Wei Pan$^{1,2,3}$}

\affiliation{$^1$Hefei National Laboratory for Physical Sciences at the Microscale and Department of Modern Physics, University of Science and Technology of China, Hefei 230026, China}
\affiliation{$^2$Shanghai Branch, CAS Center for Excellence in Quantum Information and Quantum Physics, University of Science and Technology of China, Shanghai 201315, China}
\affiliation{$^3$Shanghai Research Center for Quantum Sciences, Shanghai 201315, China}
\affiliation{$^4$State Key Laboratory of Surface Physics, Institute of Nanoelectronics and Quantum Computing,and Department of Physics, Fudan University, Shanghai 200433, China}
\affiliation{$^5$Collaborative Innovation Center of Advanced Microstructures, Nanjing 210093, China}
\affiliation{$^6$MinJiang Collaborative Center for Theoretical Physics, College of Physics and Electronic Information Engineering, Minjiang University, Fuzhou 350108, China}

\begin{abstract}
  We report an experimental study of quench dynamics across the
  superfluid transition temperature $T_{\rm c}$ in a strongly interacting
  Fermi gas by ramping down the trapping potential. The nonzero
  quasi-condensate number $N_0$ at temperature significantly above $T_{\rm c}$
  in the unitary and the BEC regimes reveals the pseudogap
  physics. Below $T_{\rm c}$, a rapid growth of $N_0$ is accompanied by
  spontaneous generation of tens of vortices. We observe a
  power law scaling of the vortex density versus the quasi-condensate
  formation time, consistent with the Kibble-Zurek theory. Our work
  provides an example of studying emerged many-body physics by quench
  dynamics and paves the way for studying the quantum turbulence in a
  strongly interacting Fermi gas.
\end{abstract}

\date{\today}


\maketitle

In pursuit of correlated quantum physics in strongly interacting Fermi gases,
great efforts have been devoted to studying equilibrium phases and
transitions~\cite{Bloch2008RoMP,Giorgini2008RoMP,Chin2010RoMP,Esslinger2010ARoCMP,Chen2014FoP,Mueller2017RoPiP}. This has shed light on the understanding of high-$T_{\rm c}$
superconductivity~\cite{Timusk1999RoPiP,Lee2006RoMP,Lee2007RoPiP} and the modeling of equation of states of dense neutron stars~\cite{GORDON1969N}.
Of equal importance would be to probe the non-equilibrium dynamics during a temperature quench across the superfluid transition temperature $T_c$, where the superfluid growth is closely connected to the
generation of spontaneous vortices.

For bosonic systems, the quench dynamics has been intensively studied~\cite{Sadler2006N,Weiler2008N,Lamporesi2013NP,Corman2014PRL,Navon2015S,Chomaz2015NC,Anquez2016PRL} and can be well described by the Kibble-Zurek (KZ) theory~\cite{Kibble1976JoPAMaG,Zurek1985N}. Very recently, obervation of the KZ scaling was also reported in Fermi gaes~\cite{Ko2019NP}. However, it is expected that the quench dynamics of strongly interacting Fermi systems should possess much richer physics due to the complexity of fermionic superfluid formation. Fermionic atoms have to  pair into bosonic degrees of freedom, Cooper pairs or bound molecules, for the formation of a superfluid. In addition to the transition temperature $T_{\rm c}$, there exists another characteristic temperature $T^*$, characterizing the onset of pair formation. In the weak coupling BCS limit, pair formation and pair condensation occur essentially at the same temperature, leading to a rapid growth of superfluid fraction as the temperature $T$ is lowered across $T_{\rm c}$. However, as the pairing strength increases, these two temperatures become distinct, and pairs can preform far above $T_{\rm c}$. This leads to a pseudogap in the fermionic excitation spectrum. At the same time, isolated superfluid islands having random relative phases may also appear above $T_{\rm c}$. As the temperature decreases, they may merge to generate vortices spontaneously. Finally, superfluidity with global phase coherence is gradually established with the annihilation of these vortices and anti-vortices. Therefore, the quench dynamics offers a great opportunity for understanding the interplay among the formation of bosonic pairs, superfluid phase coherence, and spontaneous vortices.

Here, we report an experimental study of real-time dynamics of superfluid growth and spontaneous vortex formation in a strongly interacting Fermi gas of $^6$Li atoms. We rapidly ramp down the potential of the oblate optical trap so that the system is effectively thermally quenched across the superfluid transition. For a given ramping time, the quasi-condensate number $N_0$ (consisting of bosonic pairs in the vicinity of zero momentum) is recorded in real time, while the spontaneously generated vortex density $\rho_v$ is measured upon $N_0$ reaching saturation. The observed growth dynamics of $N_0$ agree with  calculations based on the paring fluctuation theory~\cite{ChenPRL1998,ChenPhysRep2005}, by assuming that the system temperature $T$ decreases linearly with the evolution time $t$ during the ramp. The pseudogap physics is clearly revealed by the evolution of the growth dynamics of $N_0$ throughout the BCS-BEC crossover. At unitarity, for normal quenches with ramping time $t_{\rm ramp} \ge 600$~ms, the quasi-condensate formation time $t_\text{f}$ linearly increases with $t_{\rm ramp}$ and the growth dynamics of $N_0$ nicely collapse onto a single universal curve. In contrast, for  fast quenches with  $t_{\rm ramp} \leq 400$~ms, $t_\text{f}$ drops significantly as $t_{\rm ramp}$ increases, and the growth curves of $N_0$ exhibit a significant deviation from the collapse, both of which hint the breakdown of the quasi-equilibrium condition. Furthermore, by using $t_\text{f}$ as the quench time, which is less sensitive to the pseudogap physics, a power-law scaling of $\rho_v$ versus $t_\text{f}$ is observed for normal quenches, and the extracted critical exponent agrees quantitatively with that predicted by the KZ theory.

\begin{figure}
\centering
\includegraphics[width=\columnwidth]{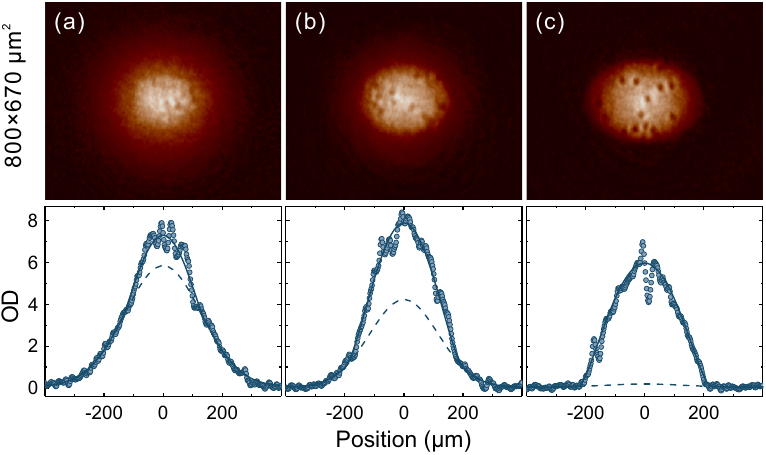}
\caption{Illustration of the real-time formation dynamics of
  quasi-condensate and spontaneous vortices, at unitarity (832~G) with
  $t_{\rm ramp} = 600$~ms.  The top row shows the absorption images of
  the cloud after 10~ms time-of-flight (TOF) at $t=520$~ms, 580~ms,
  and 700~ms, respectively, from left to right. Here, $t$ is defined
  as the evolution time of the system, i.e., $t=0$ marks the start of
  the quench. Plots in the bottom row are central line cuts of the
  column density distribution. The solid lines are the fits with a
  Gaussian plus Thomas-Fermi distribution, and the dashed lines
  indicate the Gaussian part alone. The fitting yields the
  quasi-condensate fraction $N_0(t)/N_{\rm 0, sat}\approx 0.1, 0.5$,
  and 1 (from left to right), corresponding to the initial increase,
  rapid growth and saturation stages of the quasi-condensate number,
  respectively.  }
\label{fig1}
\end{figure}

The main experimental setup and method for preparing the $^6$Li superfluid
have been described in our previous works~\cite{Yao2016PRL}.  We start
by preparing a spin-balanced mixture of $1\times 10^7$ atoms at 832.18~G in an elliptical optical dipole trap ($1/e^2$ radius 200~$\mu$m and 48~$\mu$m (in the gravity direction)). Further evaporative cooling is performed by ramping down the trap depth and holding for 3~s, yielding a superfluid of $3.9(1)\times 10^6$ atoms at about 0.3 $T_{\rm c}$. With a short ramping time, i.e.,  $t_{\rm ramp}$ varies from 200~ms to 1500~ms, temperature quench across the superfluid transition can be achieved, during which plenty of vortices are spontaneously generated~\cite{Weiler2008N,Lamporesi2013NP,Navon2015S}.
\begin{figure}
  \centering
  \includegraphics[width=\columnwidth]{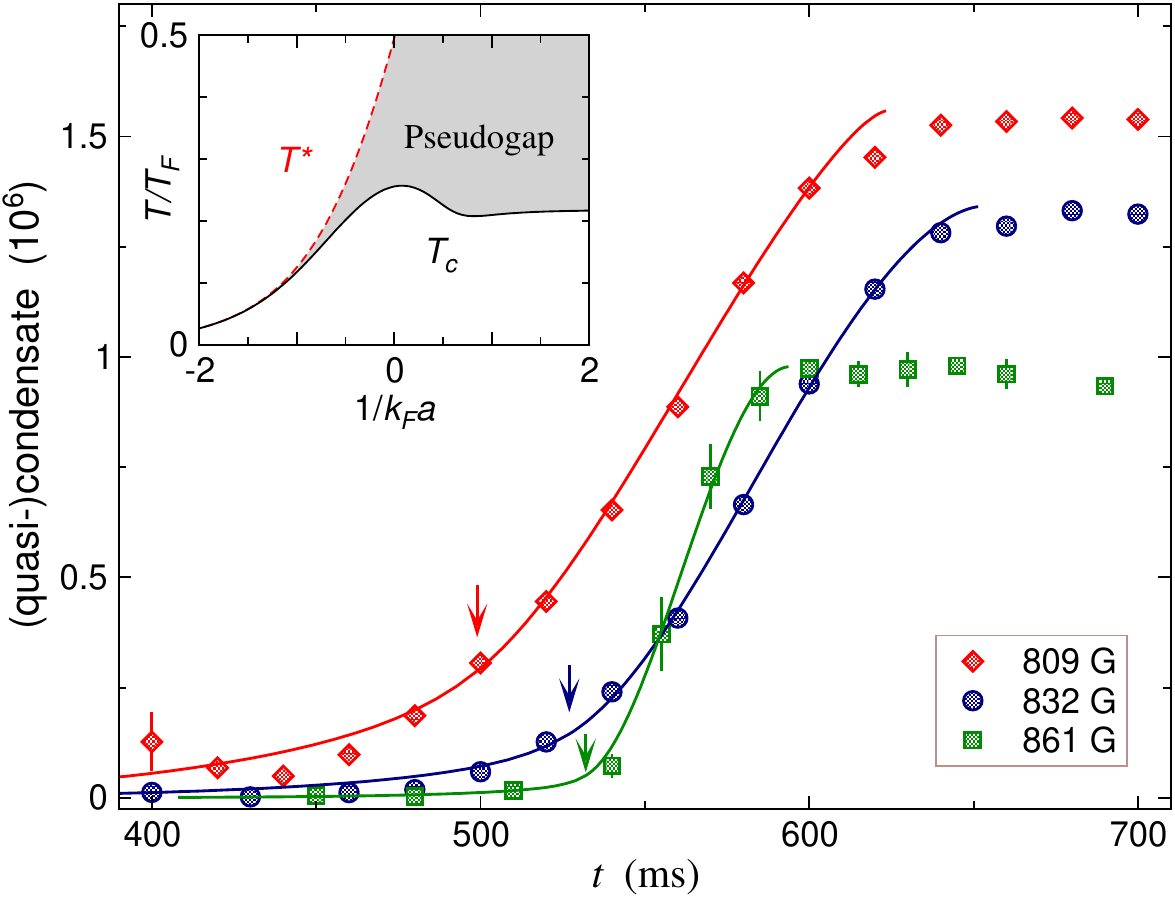}
  \caption{Real-time dynamics of quasi-condensate number $N_0$ with $t_{\rm ramp}=600$~ms, at the magnetic field of 809~G (red diamonds, BEC side), 832~G (blue circles, unitarity) and 861~G (green squares, BCS side). Each data point is an average of about 10 individual measurements with standard statistical error.  The curves represent the theoretical results calculated based on the pairing fluctuation theory, which have been re-scaled and horizontally shifted to fit the experimental data. The arrows indicate locations of superfluid transition from theory. The inset shows the phase diagram of a 3D homogeneous Fermi gas in the BCS-BEC crossover as a function of $1/k_Fa$, which manifests a pseudogap region between $T_{\rm c}$ and $T^*$. Here $k_F$ and $a$ denote the Fermi momentum and the s-wave scattering length, respectively. }
  \label{fig2}
  \end{figure}

To probe the quasi-condensate and vortices, the optical trap is suddenly switched off and the magnetic field is rapidly ramped to 720~G. After expansion for a total time of 10~ms,  strong saturation absorption imaging along the gravity direction is performed. The quasi-condensate number $N_0$ is then obtained by fitting the density profile of the cloud with a Gaussian plus Thomas-Fermi distribution. The dynamic formation of vortices is clearly visible, as shown in Fig.~\ref{fig1}. When $N_0$ is small, the vortex cores are blurred with very low contrast and are distributed in a small spatial region. As $N_0$ increases, the vortices become more visible and spread over the entire cloud. This gives a direct and vivid illustration of the evolution of superfluid coherence and the formation of spontaneous vortices. We mention that owing to the oblate trap geometry, the cloud expands rapidly in the gravity direction, resulting in a reduced imaging resolution. Nevertheless, upon saturation of $N_0$, a high contrast of  vortex cores is still achieved (see Fig.~\ref{fig1}(c)), suggesting a straight alignment of the vortex lines.

We first investigate the growth dynamics of the quasi-condensate in the BCS-BEC crossover for $t_{\rm ramp}=600$~ms. Figure~\ref{fig2} shows the quasi-condensate number $N_0$ as a function of $t$ for three typical magnetic fields of 809~G, 832~G, and 861~G. Here, $t$ is the evolution time of the system, starting at the beginning of the quench. All three $N_0$ curves seem to have a similar shape, with an initial slow increase, followed by a rapid condensate formation, and finally a nearly flat saturation.  A closer look at the growth of $N_0$ reveals the qualitative difference as the magnetic field increases. In the initial slow increase phase, $N_{\rm 0, ini}$ is clearly nonzero at 809~G (BEC) and 832~G (unitarity), while it remains nearly zero for 861~G (BCS).  During the rapid growth stage, the formation rate of the quasi-condensate monotonically increases from the BEC to the BCS regimes.

To better understand the dependence of the $N_0$ growth on the interaction strength (magnetic field), we numerically calculate the equilibrium quasi-condensate number $N^\text{th}_0$
based on the pairing fluctuation theory~\cite{Chen2005}. The pair dispersion $\Omega_\mathbf{q}\approx \hbar^2q^2/2M-\mu_\text{pair}$, or equivalently the effective pair mass $M$ and the chemical potential $\mu_{\rm pair}$, can be extracted from the pair propagator or the particle-particle scattering ${\mathbb T}$ matrix. Given the temperature, interaction strength, we are able to calculate the fermionic chemical potential $\mu$, the pairing gap $\Delta$, and the superfluid order parameter $\Delta_{sc}$  in the trap using the local density approximation. Note that the measured quasi-condensate number contains bosonic pairs with both zero and small finite momenta. Thus, we choose a small energy cutoff $\Omega_{\rm c}$, and obtain the density profile of the quasi-condensate $n_0(r)$ by summing over all the pairs with energy $\hbar^2q^2/2M< \Omega_{\rm c}$, i.e., $n_0^{}(r) = \int_{q<\Lambda}\frac{d^3q}{(2\pi)^3} \, b(\Omega_\mathbf{q}(r))$, where $b(x) = 1/(e^{x/k_BT}-1)$ is the Bose distribution function and the cutoff $\Lambda = \sqrt{2M \Omega_{\rm c}}$. Here, the energy cutoff is simply taken as $\Omega_{\rm c} = k_BT/2$, in accordance with the experimental measurements~\footnote{For qualitative comparison between experiment and theory, fine-tuning of $\Omega_{\rm c}$ is not necessary}. Finally, we  obtain the quasi-condensate number $N^{}_0=\int {d}^3 r \,n_0^{}(r)$ as a function of $T$.

To compare with the experimental growth dynamics of $N_0$, we assume a
simple linear relation between evolution time $t$ and temperature $T$
before $N_0$ saturates at very low $T$, especially during the
condensate formation stage. The theory curves are scaled in a way to
match the saturation value $N_{\rm 0,sat}$ at low $T$ and the slope at
half saturation of their experimental counterpart.  The arrows in
Fig.~\ref{fig2} indicate the superfluid transition from theory, which
correspond to a ``critical time'' $t_{\rm c}$, when the temperature crosses
$T_{\rm c}$ in the evolution of the quench dynamics. It is known that, above
$T_{\rm c}$, a pseudogap in the fermionic excitation spectrum can emerge and
bosonic pairs of fermionic atoms can already preform. The
pair-formation temperature $T^*$ depends on the atom-atom
interaction. For illustrative purpose, the phase diagram for a 3D
homogeneous Fermi gas is shown in the inset of Fig.~\ref{fig2}, where
a pseudogap region is present between $T_{\rm c}$ and $T^*$.  In general,
$T^*$ is above $T_{\rm c}$, except in the BCS limit, where the two
temperatures merge.  In the unitary and the BEC regimes, a small but
nonzero quasi-condensate already form before the critical time $t_{\rm c}$
or above $T_{\rm c}$.  Since the correlation length $\xi$ is small above
$T_{\rm c}$, the superfluid coherence is yet to be established over large
distances, and hence the growth of $N_0$ is slow. As $T$ is lowered
across $T_{\rm c}$ (or equivalently $t>t_{\rm c}$), $\xi$ can be as large as the
linear size of the system, so that $N_0$ enters a rapid-growth period
till its saturation.  In contrast, in the BCS regime, where the
pseudogap is absent, the pair formation and pair condensation roughly
occur at the same temperature. As a result, $N_0$ remains nearly zero
during the initial slow increase stage before entering an abrupt rapid
growth immediately after $t_{\rm c}$, as seen in the experimental data at
861~G. Therefore, our experiment clearly reveals the pseudogap physics
described in the theory.

\begin{figure}
\centering
\includegraphics[width=\columnwidth]{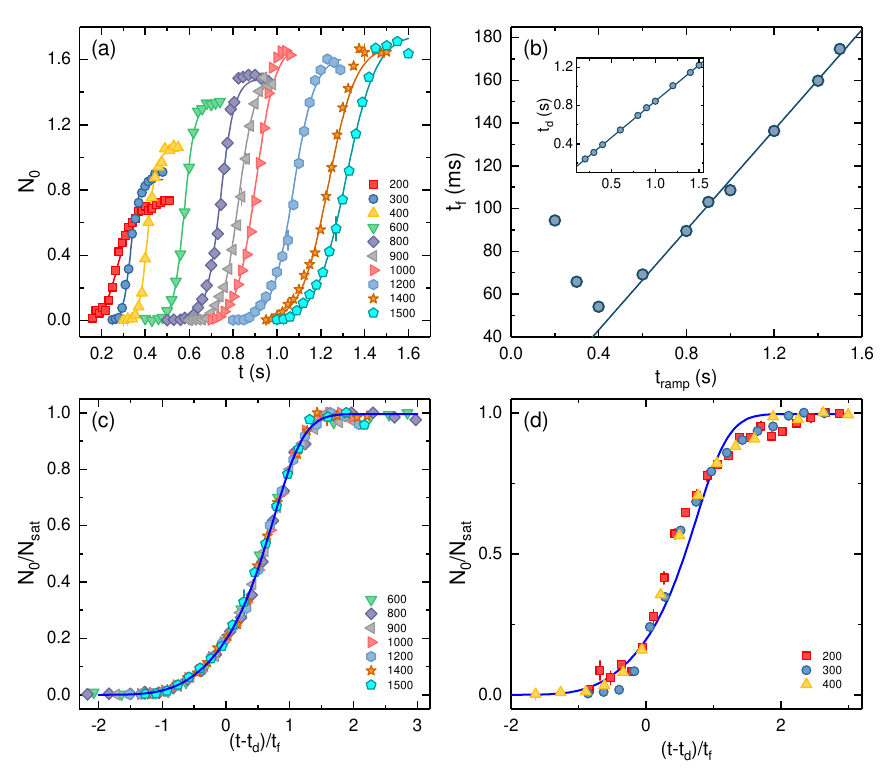}
\caption{Real-time dynamics of the quasi-condensate $N_0$ at unitarity. Each data point represents an average of about 10 individual measurements with the standard error bar. (a) Growth curves with $t_{\rm ramp}$ ranging from 200~ms to 1500~ms, where the solid curves are guides to the eye. (b) Formation time $t_\text{f}$ and delay time $t_\text{d}$ (inset) versus $t_{\rm ramp}$. Solid lines are linear fittings. (c) $N_0/N_{\rm 0, sat}$ versus $(t-t_\text{d})/t_\text{f}$, for normal quenches with $t_{\rm ramp} \geq 600$~ms. The data are fitted with a smoothing spline (solid line). (d) $N_0/N_{\rm 0, sat}$ data for fast quenches with $t_{\rm ramp} \leq 400$~ms, compared with the solid curve for normal quenches.}
\label{fig3}
\end{figure}

Next, we study the dependence of the quench dynamics on the ramping time $t_{\rm ramp}$. Shown in Fig.~\ref{fig3}(a) are the growth curves of $N_0$ at unitarity for $t_{\rm ramp}$ ranging from 200~ms to 1500~ms. As $t_{\rm ramp}$ becomes longer, the saturated quasi-condensate number $N_{\rm 0, sat}$ also increases because of the less atom loss during the evaporative cooling. It is seen that for quenches with $t_{\rm ramp} \! \ge \! 600$~ms, $N_0$ roughly reaches its saturation at the end of quench. In contrast, for quenches with $t_{\rm ramp} \! \le \! 400$~ms, the rapid formation of $N_0$ has barely started by $t=t_{\rm ramp}$, and the much suppressed $N_{\rm 0, sat}$ is not reached until a much later time. To better describe the quench dynamics, we introduce two time scales, delay time $t_\text{d}$ and formation time $t_\text{f}$, corresponding to the starting time and the duration of the rapid formation of $N_0$, respectively.  In practice, they are determined via $N_0(t_\text{d})/N_{\rm 0, sat} = 0.2$ and $N_0(t_\text{d}+t_\text{f})/N_{\rm 0, sat} = 0.8$, respectively \footnote{Note that slight variation in the cutoff percentages does not change the power law exponent in the KZ scaling in Fig. 4.
  }.  As shown in Fig.~\ref{fig3}(b), $t_\text{d}$ follows a nice linear increasing function of $t_{\rm ramp}$ for all the quenches. However, as $t_{\rm ramp}$ increases, $t_\text{f}$ first decreases until it reaches a minimum around $t_{\rm ramp} \approx 400$~ms, and then increases linearly.  Based on this observation, we classify the quenches into two types, normal and fast ones, which are separated at $t_\text{ramp}\sim 500$~ms for our system.  By plotting $N_0(t)/N_{\rm 0, sat}$ versus $(t-t_\text{d})/t_\text{f}$, we find that all experimental data for normal quenches can be well described by a single universal curve (see Fig.~\ref{fig3}(c)), while those for fast quenches exhibit a significant deviation from this curve (Fig.~\ref{fig3}(d)).

\begin{figure}
\centering
\includegraphics[width=\columnwidth]{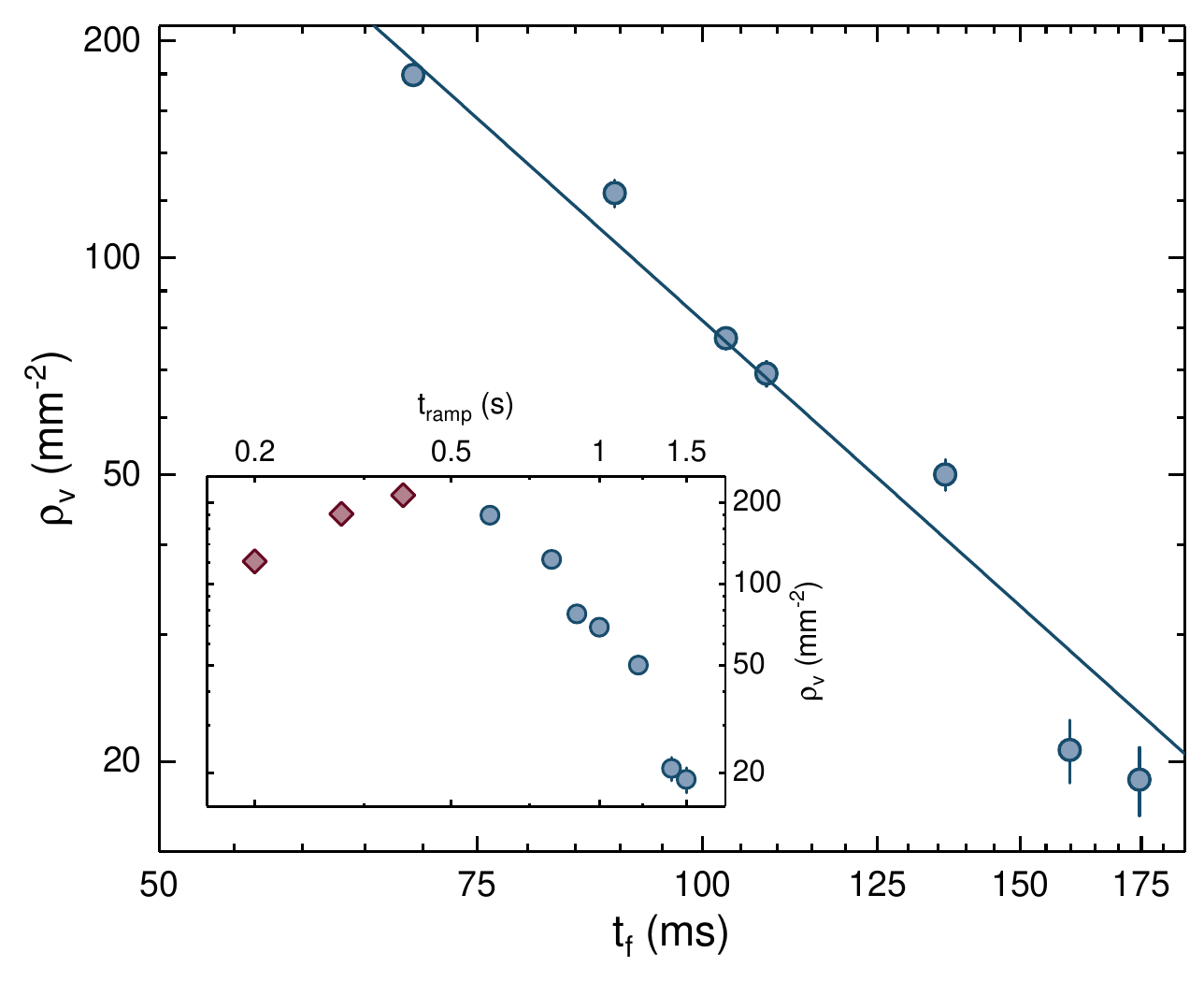}
\caption{Log-log plot of vortex density $\rho_v$ versus $t_\text{f}$.
The error bar for each point represents the standard statistical error over 30 independent measurements. The solid straight line is the power-law fitting curve based on the KZ theory. The inset shows $\rho_v$ as a function of $t_{\rm ramp}$, where blue circles and red squares denote normal and fast quenches, respectively.}
\label{fig4}
\end{figure}

We now study the spontaneous generation of vortices in the quench dynamics, by measuring the vortex density $\rho_v$ at the saturation of quasi-condensate for each $t_{\rm ramp}$. It is known that near the superfluid transition $T_{\rm c}$, a diverging correlation length develops as $\xi \! \sim \! |T-T_{\rm c}|^{-\nu}$ and the relaxation time diverges as $\tau \! \sim \! \xi^{z}$, with $\nu$ and $z$ being the static and dynamic critical exponents, respectively~\cite{Kibble1976JoPAMaG,Zurek1985N}. Under the condition that the temperature $T$  varies linearly with time near $T_{\rm c}$, the KZ theory predicts that $\rho_{v}$ decays algebraically  with the quench rate $1/\tau_{\rm Q}$ as $\rho_{v} \! \sim \! \tau_{\rm Q}^{-\alpha_{\rm KZ}}$, where the exponent $\alpha_{\rm KZ}$ is determined by $\nu$ and $z$. Experimentally, the measurement of temperature evolution in quench dynamics is a great challenge for strongly interacting Fermi gases. In previous  studies, $\tau_{\rm Q}\sim t_{\rm ramp}$ has been reported~\cite{Lamporesi2013NP, Donadello2016PRA,Ko2019NP,Goo2021arXiv}, and thus we first attempt to plot $\rho_v$ versus $t_{\rm ramp}$. As shown in the inset of Fig.~\ref{fig4}, an approximate power-law decay is observed for normal quenches, while for fast quenches the $t_{\rm ramp}$ dependence of $\rho_v$ clearly deviates from the KZ scaling.

To understand the normal and fast quenches better, we revisit the relaxation dynamics of an out-of-equilibrium system. For a superfluid, there are two types of excitations, i.e., low-energy density waves and high-energy vortices. Typically, the relaxation of low energy modes is much faster than the annihilation of vortex and anti-vortex pairs. The quasi-equilibrium condition is assumed that, at each evolution time, the low energy modes have been sufficiently relaxed while the vortices remain excited. For normal quenches, the quasi-equilibrium condition is supported by the observations that the formation time $t_\text{f}$ of the quasi-condensate increases linearly with $t_{\rm ramp}$ (see Fig.~\ref{fig3}(b), $N_0$ almost reaches $N_{\rm 0,sat}$ at the end of the quench, and that the saturated vortex density $\rho_v$ decays algebraically.  On the other hand, the unusual $t_{\rm ramp}$ dependence of $t_\text{f}$ and $\rho_v$, as well as the barely started growth of $N_0$ by $t=t_{\rm ramp}$, suggest that the quasi-equilibrium condition is broken for fast quenches.

In Fig.~\ref{fig4}, the data points of $\rho_v$ versus $t_\text{f}$ for normal quenches agree well with a power-law scaling. Indeed, $t_\text{f}$ reflects the linear growth period of $N_0$ and the linear decrease of temperature with time. Unlike $t_\text{ramp}$, it is insensitive to the  (somewhat arbitrary) initial temperature of the system ($T$ at $t=0$) as well as the complications caused by the pair formation process during the slow incubation stage. Therefore, it is inversely proportional to the  quench rate near $T_{\rm c}$ and thus may naturally play the role of $\tau_Q$ in the KZ theory. Fitting the experimental data  with a power-law function $\rho_v \sim t_\text{f} ^{-\alpha_{\rm KZ}}$, we obtain  the KZ exponent $\alpha_{\rm KZ}=2.25(17)$. In a 3D harmonic trap, the KZ exponent has been predicted to be $\alpha_{\rm KZ} = 2 (1+2\nu)/(1+\nu z)$~\cite{Zurek2009PRL,Campo2011NJoP}. According to the F model, $\nu=2/3$ and $z=3/2$ for a 3D system~\cite{Hohenberg1977RoMP}, which yields $\alpha_{\rm KZ} =7/3$. Our experimental result is in quantitative agreement with this theoretical value, demonstrating the validity of using $t_\text{f}$ to characterize the quench rate for normal quenches.

In conclusion, we have studied the quench dynamics of a strongly interacting atomic Fermi gas by ramping down the trapping potential. Our experiment directly demonstrates the interplay between the real-time dynamics of quasi-condensate growth and spontaneous vortex formation. Comparison between theoretical calculations and experimental data reveals the pseudogap physics, which leads to significant differences in the growth dynamics of quasi-condensate between the BEC and BCS regimes. We find that the quench processes can be classified into
normal and  fast quenches. The unusual non-monotonic $t_\text{ramp}$ dependence of the quasi-condensate formation time $t_\text{f}$ and the vortex density $\rho_v$ suggests that the quasi-equilibrium condition is broken during the fast quench processes. For normal quenches, by using $t_\text{f}$ to characterize the quench time of the system, the KZ scaling of strongly interacting Fermi gas is observed and the extracted KZ exponent agrees well with the theoretical prediction. Our work may serve as a starting point for exploring rich quantum phenomena of quasi-2D vortices, such as Berezinskii-Kosterlitz-Thouless physics in a quasi-2D trap \cite{Berezinskii,*Kosterlitz}, holographic liquids~\cite{Chesler2013}, and quantum turbulence~\cite{Tsubota2009CP, Barenghi2014PotNAoS, Bulgac2016JoPBAMaOP}.

This work is supported by the National Key R\&D Program of China (Grant Nos. 2018YFA0306501, 2017YFA0304204, 2016YFA0301604), NSFC of China (Grant Nos. 11874340, 11425417, 11774067, 11774309, 11625522, 11934002), the Chinese Academy of Sciences (CAS), the Anhui Initiative in Quantum Information Technologies, the Shanghai Municipal Science and Technology Major Project (Grant No.2019SHZDZX01).

\end{document}